# Mixing Algorithm for Extending the Tiers of the Unapparent Information Send through the Audio Streams

Sachith Dassanayaka

Department of Mathematics & Computer Science, Wittenberg University, Ohio, 45501, USA
dassanayakas@wittenberg.edu

**Abstract** – Usage of the fast development of real-life digital applications in modern technology should guarantee novel and efficient way-outs of their protection. Encryption facilitates the data hiding. With the express development of technology, people tend to figure out a method that is capable of hiding a message and the survival of the message. Secrecy and efficiency can be obtained through steganographic involvement, a novel approach, along with multipurpose audio streams. Generally, steganography advantages are not used among industry and learners even though it is extensively discussed in the present information world. Information hiding in audio files is exclusively inspiring due to the compassion of the Human Auditory System (HAS). The proposed resolution supports Advance Encryption Standard (AES)256 key encryption and tolerates all existing audio file types as the container. This paper analyzes and proposes a way out according to the performance based on robustness, security, and hiding capacity. Furthermore, a survey of audio steganography applications, as well as a proposed resolution, is discussed in this paper.

## 1. Introduction

The professionals and researchers focus on protecting information due to the popularity of the Internet among the public and the availability of private data. The importance of confidentiality of sensitive information in the military sector has persisted to this day, but it is also essential in other areas. So, areas such as information technology and computer networks that deal with computer systems require powerful resolutions for their survival. Watermarking, cryptography, and steganography are the prominent areas of data protection. In watermarking, ownership and copyright information are hidden inside the cover object, while cryptography techniques are based on the jumbled content of a message interpreted by unauthorized people. However, cryptography and watermarking skills are outstanding for supporting data security; a sharp interest in exploring complex, fresh approaches has been the focus of ongoing research. The following Table I expresses the dissimilarities and the similarities between watermarking, cryptography, and steganography. Steganography is to accurately transport hidden information secretly, not merely to make its existence ambiguous.

Table I
Summary of Digital Data Security

| Technique | Cryptography | Information Hiding | |
| --- | --- | --- | --- |
| | | Steganography | Watermarking |
| Application | Secret Information | Secret Communication | Copyright |
| Requirement | Robustness | Unpredictable | Robustness |
| Principle | Text + Key + Encryption = Cypher Text | Secret Message + Key + Cover object + Embedding = Embedded File | Signature + Key + Cover object + Encoding = Marked File |
| Visibility | Visible | Invisible and Silent | Visible/ Invisible |

New directions of steganography approaches emerge to protect the secret data. However, through steganography techniques, data security can achieve better secrecy through modern, predictable security approaches. Currently, steganography techniques use the robust characteristics of digital media as the cover object to hide information. These containers can be any file type, such as audio, video, image, or text document. The core workflow of digital audio steganography can be illustrated in Figure 1. The sender embeds secret information inside a digital audio cover object and a secret key, concealing the hidden message's existence [1]. At the receiver's end, the receiver extracts the secret message from the digital audio file using the same secret key to embed the information. Currently, steganography approaches exploit natural boundaries in human auditory and visual sensitivities.

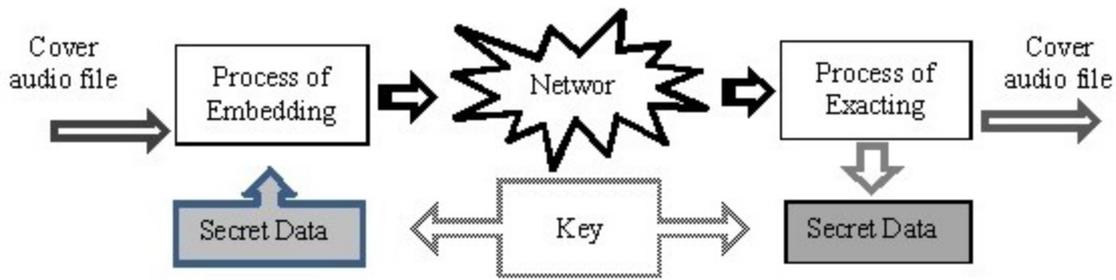

Figure 1: Workflow of Audio Steganography

Steganography techniques in image, video, and audio rely on exploiting the limitations of human perception. For visual media, the human eye struggles to detect subtle luminance variations, allowing for the concealment of information. In audio steganography, the Human Auditory System (HAS) is leveraged to hide data within sound files [2, 3]. The effectiveness of audio steganographic methods depends on various factors, including the specific application and transmission environment. There is often a trade-off between robustness and data hiding capacity, with increased robustness typically resulting in reduced capacity [1]. These factors and the chosen method influence the quality of the final steganographic output. This research explores a novel algorithm in audio steganography, focusing on the use of digital audio files as carriers for concealed communication, and examines the latest techniques in this field.

In the digital age, fake review generators are adopting sophisticated techniques from the world of steganography, the art of hiding information in plain sight. Just as traditional steganography conceals messages within innocuous files, advanced fake review systems are now capable of embedding hidden data within seemingly authentic product evaluations [4, 5]. These systems employ linguistic steganography, manipulating subtle aspects of language such as word choice, sentence structure, and punctuation patterns to encode additional layers of meaning invisible to casual readers. This technology presents a significant challenge to detection systems, as reviews generated using these methods can appear entirely genuine on the surface while carrying concealed information. As detection algorithms evolve, so do the steganographic techniques, creating an ongoing arms race in the realm of online deception. The implications of this technology extend beyond simple product reviews, potentially enabling covert communication networks or allowing marketers to embed tracking data within organic-looking endorsements. As this field continues to advance, it becomes increasingly crucial for both technological and regulatory approaches to keep pace, ensuring the integrity of digital discourse in an era where the very words we read online may harbor hidden meanings.

The rise of misinformation in the digital landscape further complicates this issue, as it intersects with the steganographic techniques used in fake reviews [6, 7]. Misinformation, defined as false or inaccurate information spread unintentionally, can be weaponized when combined with hidden messages in reviews. For instance, a seemingly innocent product review might contain both misinformation about the product's features and steganographically encoded data, creating a dual layer of deception. This combination can erode consumer trust in brands, as users become increasingly aware of the pervasiveness of misinformation in online spaces. The challenge for consumers and brands alike is not only to discern genuine reviews from fake ones but also to navigate the complex web of intentional and unintentional misinformation that may be concealed within them. As this trend continues, it becomes crucial for both technological and regulatory approaches to keep pace, ensuring the integrity of digital discourse in an era where the very words we read online may harbor hidden meanings and unintended falsehoods.

## 2. Background

Steganography, derived from the Greek words "steganos" (secret or covered) and "graphia" (writing), is the art and science of concealing information within seemingly innocuous carriers [8]. The concept has a rich history, with the first documented techniques appearing in Johannes Trithemius' 1499 book "Steganographia". In modern times, digital audio files have become a significant medium for steganographic practices due to their ubiquitous nature and the complexity of the human auditory system.

The human ear's intricate mechanisms provide natural opportunities for information hiding [2]. For instance, loud sounds can effectively mask softer ones, and listeners often overlook subtle environmental audio distortions. Researchers have capitalized on these auditory properties to develop sophisticated steganographic techniques for digital audio streams. The effectiveness of these methods is governed by several key properties, including hiding rate, imperceptibility, and resilience to various signal processing operations such as amplification, filtering, re-sampling, noise addition, encoding, and transcoding [9-14]. Successful steganography in audio must strike a delicate balance between these factors to ensure that the hidden information remains undetectable to human listeners and automated steganalysis tools, which are generally less developed than steganographic techniques [15].

### 2.1 Audio Steganography Approaches

Steganography techniques worry about security, forcefulness, and hiding capacity. The encoding process limits the audio's size by removing useless data. Three unique embed categories can be used in audio steganographic resolutions: in-encoder, pre-encoder, and post-encoder embedding [16].

#### 2.1.1 In-encoder embedding

This technique tolerates the robustness of embedded data. The process of embedding data exists within the codec. However, hidden data will be modified if the encoded parameters cross a network such as GSM. The same modification can occur if the Radio Access Network activates a sound improvement algorithm [16].

#### 2.1.2 Pre-encoder embedding

This technique, which embeds data before the process of encoding, applies to frequency and time. This method does not guarantee the reliability of the hidden data through the network due to adding noise, such as white Gaussian noise (WGN). A noise-free background can be attained by using a high embedding data rate [16].

#### 2.1.3 Post-encoder embedding

Data are embedded after the encoder and extracted before traversing the decoder side. Modification of the bits should be minimized to avoid the perceptibility of the embedded data. The accuracy of the extracted data can be guaranteed through this embedding technique. The encoding process does not affect the extraction of the hidden message [16].

## 3. Existing Strengths

Prominent judgments of this research are based on specific philosophies in computer science as well as statistics. Java programming language has been used for application development. However, the following areas can be noticed as the leading foundation of this research.

### 3.1 Noise

Audio segmental signal-to-noise ratio (SNR) is used to explain the average value of all SNR values in audio [15]. Embedded data in the audio $s_c(m, n)$ induces the distortion value which represents the segmental SNR. The SNR value can be presented as follows.

$$\text{SNR}_{dB} = 10\log_{10}\left(\frac{\sum_{n=1}^{N}|s_c(m,n)|^2}{\sum_{n=1}^{N}|s_c(m,n)-s_s(m,n)|^2}\right) \quad (1)$$

The embedded audio signal represents $s_s(m,n)$ where m=1,..., M and n=1,...N. M and N are the number of frames per milliseconds (ms) and the number of samples for each frame, respectively [17-26].

### 3.2 Sampling Rate

Capture audio covering the entire 20–20,000 Hz range of human hearing, such as when recording music or many types of acoustic events, and audio waveforms are typically sampled at 44.1 kHz, 48 kHz, 88.2 kHz, or

96 kHz[27]. The sample rate $f_s$, is the average number of samples in one second. In other words, it says to take samples per second. Let t be the time in seconds [28]. So that we can say,
$f_s = 1/t$.
The approximately double-rate condition is a consequence of the Nyquist theorem [29]. Only derives the series coefficients for the case $f_s = 2B$. Virtually citing Shannon's original paper:

Let $X(\omega)$ be the spectrum of $x(t)$, then

$$X(\omega) = \frac{1}{2\pi}\int_{-\infty}^{\infty} X(\omega)e^{i\omega t}\, d(\omega)$$

$$= \frac{1}{2\pi}\int_{-2\pi B}^{2\pi B} X(\omega)e^{i\omega t}\, d(\omega) \quad (2)$$

Since $X(\omega)$ is assumed to be zero outside the band $|\frac{\omega}{2\pi}| < B$. If we let

$$t = \frac{n}{2B} \quad (3)$$

where n is any positive or negative integer, then

$$x\left(\frac{n}{2B}\right) = \frac{1}{2\pi}\int_{-2\pi B}^{2\pi B} X(\omega)e^{i\omega \frac{n}{2B}}\, d(\omega) \quad (4)$$

### 3.3 Cryptography

Cryptography, initially synonymous with encryption, has evolved significantly over time. At its core, it involves the transformation of intelligible information into an apparently meaningless form. This research focuses on the Advanced Encryption Standard (AES), a widely adopted and highly regarded encryption method that has gained acceptance by the U.S. government and has become a cornerstone in the field of information security [30].

AES, which superseded the Data Encryption Standard (DES) of 1977, is a symmetric key algorithm. This means the same key is used for encryption and decryption processes, necessitating secure key exchange between communicating parties. Unlike its predecessor, AES is not a Feistel cipher; instead, it simultaneously operates on the entire input block. The algorithm supports block sizes of 128 bits (with options for 192 or 256 bits) and offers varying key lengths of 128, 192, or 256 bits. Depending on the chosen configuration, AES performs 10, 12, or 14 rounds of encryption, utilizing 44, 52, or 60 sub-keys, each 32 bits in length. This flexibility in configuration allows for a balance between security strength and computational efficiency, making AES adaptable to a wide range of applications in modern cryptography.

### 3.4 Least significant bit

The Least Significant Bit (LSB) technique is a widely adopted steganographic method in real-world applications, applicable to both image and audio cover objects [31]. Traditionally, this method involves modifying the least significant (8th) bit of each byte in the data stream to embed hidden information. However, this research explores a modified LSB approach that targets the 7th bit position instead of the conventional 8th bit. This slight adjustment can potentially offer a balance between increased data capacity and maintaining imperceptibility, as alterations to the 7th bit may still be subtle enough to evade detection while allowing for more substantial data embedding [32]. By shifting the focus to a less prominent bit position, this modified technique aims to enhance the robustness of the steganographic process while preserving the essence of the LSB method's simplicity and effectiveness.

## 4. Proposed way-out

The resolution of audio steganography encompasses various approaches, all operating within the In-encoder Embedding architecture framework. This technique is adaptable to the diverse array of audio file formats prevalent in today's digital landscape, each possessing it's unique characteristics. The initial step in this process involves identifying the specific audio file format of the carrier, which is crucial for determining the header size and extracting the relevant header information. The header size varies depending on the audio format, and it's important to note that this section must remain unaltered during the steganographic process. By preserving the integrity of the header, the steganographic method ensures that the modified audio file maintains its original format and playability while still allowing for the embedding of hidden data in the subsequent sections of the file.

### 4.1 Process of Embedding

Hiding information may represent any kind of file type, and those files are called messages. These messages can be any file type, including audio and image file styles. The system calculates the hidden message size and assigns it to 4 bytes. Then, the system loads the audio file into the buffer and reads the header bytes. So that the system can identify the format of the audio file that is used as the carrier. Then, the algorithm calculates the audio file size in bytes. First, H bytes, let's say the header bytes, includes all the audio file information. So, we are not allowed to do any modification inside the first H bytes. Again, the system requires one byte to write the hide type during the hidden process. Users are able to select the hiding type according to their hide file and carrier file. Qualities are original and Excessive. Next, 8 bytes or 4 bytes are required to represent the file format of the message according to the above-mentioned hiding type because the type of the message array consists of one byte. Furthermore, another 32 bytes, or 16 bytes, require the mention of the message or document size, which the user will hide under Regular type and Excessive type, respectively. The number of bytes that represent the audio header, the type of hide after the process, the file format of the message, and the message size are called reserved bytes. The process inside the system can be categorized according to the size of the hide file and carrier file. Let's say (audio bytes – reserved bytes) = K, where K is a positive integer value. If K/4 is greater than the size of the hidden message, then only excessive type can be chosen. If K/8 is greater than the size of the hidden message, then Regular type can be chosen as well as Excessive type. The system then leads the user to select the best choice. At the initial point of the process, systems encrypt the message using AES256 with the user's given secret key. The encryption process does not change the size of the message. Load the encrypted message into the system memory and divide the encrypted message byte stream into two sub-byte streams, say m1 and m2 byte stream. If the encrypted message byte stream size is even, then m1 and m2 are equal. Otherwise, m1 is greater than the m2 by one. While writing the message bits inside audio bits, if the corresponding audio bit position represents the same bit value that compares with the message bit, it keeps that audio bit as it is. Otherwise, write the message bit value. According to the user's selected hide method, the processes can be explained as follows:

### 4.1.1 Regular Type Embed

Skip the first H bytes from the loaded audio file into the buffer and write the type (Regular represents binary 1) in $H+1^{th}$ audio byte using LSB. Write the file type of the message in the next 8 bytes referring to the same technique. Then, write the encrypted message size in the next 32 bytes using LSB so that the system passes (H+41) audio bytes. Now, write all message bits of m1, using LSB in each even audio byte position from $H+42^{th}$ audio byte position. At the same time, write all message bits of m2, each odd audio byte positions from $H+43^{th}$ audio byte. Then, write the audio byte; each two audio bytes stream into the given location with default parameter values.

### 4.1.2 Excessive Type Embed

Skip the first H bytes from the loaded audio file into the buffer and write the type (Excessive represent binary 0) in $H+1^{th}$ audio byte using LSB. Writes first 4 bits of message type next to each 4 bytes using LSB. At the same time, write the rest of the 4 bits in each $7^{th}$-bit position of the same audio bytes. Then, write the encrypted message size in the next 16 bytes as follows. Encrypted message size has been assigned to 4 bytes (i.e. 32 bits). Write the first 16 bits in every 16 bytes using LSB and, simultaneously, write the second 16 bits in each $7^{th}$-bit position of the same audio bytes. Now, the system has passed $H+20^{th}$ audio bytes. So, write all message bits of m1, using LSB in each audio byte from the $H+21^{th}$ audio byte position. At the same time, write all message bits of m2 in each $7^{th}$ bit of audio bytes from the $H+21^{th}$ audio byte position. Then, write the audio byte stream into the given location with default parameter values.

## 4.2 Process of Extraction

The retrieval process is the reverse of the encoding process. When a user needs to extract a hidden message from an audio file, they must enter the same secret key that was used for encryption during the encoding process. Then, the system loads the audio stream into a buffer and reads the first H bytes to identify the audio file type. After that, it reads the $H+1^{th}$ byte to determine the used encoding method. Finally, the retrieval activity begins according to the appropriate procedures.

### 4.2.1 Regular Type Extract

Systematically retrieve the 8th bits from each next 8 bytes, then create a byte array from these bits. This systematic approach ensures the system comprehends the format of the hidden message. Repeat this process, retrieving the 8th bits from each next 32 bytes, and creating a byte array. Then, read the byte array and extract the encrypted message size using the bit shift technique in an object-oriented programming language.

If the size of the encrypted message, m, is even, retrieve the 8th audio bit from each even audio byte position from the beginning ($H+42^{th}$ audio byte) to the 8m audio byte position. Assign these bits to a byte array called m1. Simultaneously, retrieve the 8th audio bit from each odd audio byte position from the $H+43^{th}$ audio byte to the 8(m+1) audio byte position and assign them to a byte array called m2.

If m is odd, then retrieve the 8th audio bit from each even audio byte positions from the beginning where, $H+42^{th}$ audio byte to 8(m+1) audio byte position and assign them to a byte array called m1. At the same time, retrieve the $8^{th}$ audio bit from each odd audio byte positions from the $H+43^{th}$ audio byte to 8m audio byte position and assign them to a byte array called m2.
Now, the user plays a crucial role as they concatenate both m1 and m2 and decrypt the full message using their secret key. This key step, along with the extraction of the message, is then written into the user's specified location, providing them with the extracted details of the message type.

### 4.2.2 Excessive Type Extract

Retrieve the $8^{th}$ bits from each next 4 bytes and 7th bits from the same bytes separately in a systematic manner. Then, create a byte array from these retrieved bits, ensuring the system understands the format of the hidden message. Retrieve the $8^{th}$ bit from each of the next 16 bytes and create a byte array from these bits, referred to as c1. Simultaneously, the $7^{th}$ bit from each of the 16 bytes will be retrieved, and a byte array will be created from these bits, referred to as c2. Concatenate both c1 and c2, and read the full byte array to retrieve the encrypted message size using bit shift technique in an object-oriented programming language.

If m is even, where m is the size of the encrypted message, then retrieve the 8th audio bit for each byte from the beginning, specifically from the $H+22^{th}$ audio byte to the 4m audio byte position. These bits are assigned to a byte array called m1. Simultaneously, retrieve the $7^{th}$ audio bit for each byte from the same range and assign them to a byte array called m2. This step is crucial as it helps in understanding the decryption process. If m is odd, then retrieve the $8^{th}$ audio bit for each byte from the beginning where $H+22^{th}$ audio byte to (4m+8) audio byte position and assign them to a byte array called m1. At the same time, retrieve the $7^{th}$ audio bit for each byte from the $H+22^{th}$ audio byte to the 4m audio byte position and assign them to a byte array called m2.

Now, concatenate both m1 and m2 and decrypt the full message using the user's secret key, emphasizing the user's integral role in the process. At the end, the extracted message is written into the user's specified location, along with the extracted details of the message type.

A unique feature of this system is that when a user decides to remove a message, they initiate a process that requires the secret key used to encrypt the message. The system then reads the $8^{th}$-bit position of the $H+1^{th}$ byte location and picks up the value (binary 0 or 1) to which the hidden message belongs.

If the value is 1, then it skips the next 8 bytes and retrieves the 8th bit from each of the next 32 bytes. Then, it creates a byte array from the retrieved bits. Then, it reads the byte array and retrieves the encrypted message size using the bits shift technique. Let's say the retrieved message size is y. Replace the $8^{th}$ position bit of the audio byte stream to value 0 from $H+10^{th}$ to $H+41^{th}$ audio byte and again from the $86^{th}$ audio byte to 16y. If the value is 0, the system skips the next 4 bytes and retrieves the 8th bit from each next 16 bytes, creating a byte array called c1. Simultaneously, it retrieves the 7th bit from the same 16 bytes, creating a byte array called c2. The system then concatenates both c1 and c2, reads the full byte array, and retrieves the encrypted message size using the bit shift technique. Let's say the retrieved message size is z. The system replaces the 8th position bit of the audio byte stream to 0 from $H+6^{th}$ to $H+21^{th}$ audio byte and again from $H+22^{th}$ audio byte to 2z. The system does not replace $7^{th}$ bits because additional bit replacement would introduce more noise into the carrier file.

Finally, the system replaces the $8^{th}$ position bit of the $H+1^{th}$ audio byte stream with its opposite value and overwrites the existing audio stream. Crucially, users have the power to choose the "remove message" option at any stage of the process, whether they're retrieving the message or at a later time.

Both encoding and decoding can be summarized as follows:

Let say,
Audio bytes = M
Encrypted Message bytes = m, so that message bits = 8m
Audio header = H bytes
Bytes to represent the message size for Regular type = 32 bytes
Bytes to represent the message size for Excessive type = 16 bytes
Bytes to represent the file type of the message for Regular type = 8 bytes
Bytes to represent the file type of the message for Excessive type = 4 bytes
Bytes to represent the type = 1 byte
Total reduction from the audio bytes for Regular type = 1+ 8+ 32 bytes = 41 bytes
Total reduction from the audio bytes for Excessive type = 1+ 4+ 16 bytes = 21 bytes..

Now, let's consider,
(H+41) = R1
(H+21) = R2
m/2 = n where, n is a positive integer.

Regular type:
This can be performed only if (M-R2)/8 is greater than m.
If (M-R2)/8 > m then
    If (m%2) == 0 then
        m1=m2 = n.
        Write m1 bits using LSB for each even audio byte from audio byte position R1+1 to R1+16n. (i.e. R1+1, R1+3, R1+5, …, R1+16n)
        Write m2 bits using LSB for each odd audio byte from audio byte position R1+2 to (R1+16(n+1)). (i.e. R1+2, R1+4, R1+6, …, (R1+16(n+1)))
    Else
        m1= n+1 and m2 = n.
        Write m1 bits using LSB for each even audio byte R1+1 to (R1+16(n+2)). (i.e. R1+1, R1+3, R1+5, …, (R1+16(n+2)))
        Write m2 bits using LSB for each odd audio byte R1+2 to (R1+16(n+1)). (i.e. R1+2, R1+4, R1+6, …, (R1+16(n+1)))
Else, carrier size is not enough.

Excessive type:
This can be performed only if (M-R2)/4 is greater than m.
If (M-R2)/4 > m then
    If (m%2) == 0 then
        m1=m2 = n.
        Write m1 bits using LSB for each audio byte from audio byte position R2+1 to R2+8n.
        Write m2 bits in $7^{th}$ bit position of each audio byte from audio byte position R2+1 to R2+8n.
    Else
        m1= n+1 and m2 = n.
        Write m1 bits using LSB for each audio byte from audio byte position R2+1 to R2+ 8(n+1).
        Write m2 bits in $7^{th}$ bit position of each audio byte from audio byte position R2+1 to R2+8(n+1).
Else, carrier size is not enough.
Finally, writes the audio file.

Now, consider the process of retrieval. Let say, type = q where, q is either 0 or 1.

If (q == 1) then
    If (m%2) == 0 then
        Read bits using LSB for each two byte from audio byte position R1+1 to R+16n and assign them to A1.
        Read bits using LSB for each two byte from audio byte position R1+2 to (R1+16(n+1)) and assign them to A2.
        Concatenate A1 and A2.
    Else
        Read bits using LSB for each two byte from audio byte position R1+1 to R1+16(n+2) and assign them to A1.
        Read bits using LSB for each two byte from audio byte position R1+2 to (R1+16(n+1)) and assign them to A2.
        Concatenate A1 and A2.
Else
    If (m%2) == 0 then
    Read bits using LSB in each byte from audio byte position R2+1 to R2+8n and assign them to A1.
    Read $7^{th}$ bits position in each byte from audio byte position R2+1 to R2+8n and assign them to A2. Concatenate A1 and A2.
    Else
        Read bits using LSB in each byte from audio byte position R2+1 to R2+8(n+1) and assign them to A1.
        Read $7^{th}$ bits position in each byte from audio byte position R2+1 to R2+8(n+1) and assign them to A2.
        Concatenate A1 and A2.
        Decrypt the message.

Now, consider the process of delete the message. Reads the 8th bit of $H+1^{th}$ byte and picks up the value. Let say, value is t where, t is either 0 or 1.
If (t == 1) then
    Skips next 8 bytes and reads $8^{th}$ bit of next 32 bytes. Retrieved byte says the message size y.
    Replace LSB to value 0 for each byte from audio byte position $H+10^{th}$ to R1 and again R1+1 to R1+16y.
Else
    Skips next 4 bytes. Reads $8^{th}$ bit of next 16 bytes and assign them to array c1. At the same time reads the $7^{th}$ bit of the same bytes and assign to c1. Concatenate both c1 and c2 to one byte array and read the value. Retrieved byte says the message size z.
    Replace LSB to value 0 for each byte from audio byte position $H+6^{th}$ to R2 and again R2+1 to R2+2z.

Replace LSB to value 0 of the $H+1^{th}$ audio byte position and overwrites the audio.
This application facilitates play mode feature that users to listen the audio at any time. Users are able to listen to the music at the same time while they encode or decode the message.

### 4.3 Send and Receive Encoded File

The application assists to send audio files to target location in local area network. It is individually design for sender and receiver, who use the suggest application. For each connection perform via random socket. Furthermore, more than one connections are allowed simultaneously without any restrictions and file transferring fragment is free from the embed process.

## 5. Performance

Testing cycles focused prominent categories during the processes. All the tested areas have been summarized below.
Test 01: Size and the time duration of the original audio file and embedded audio file.

Test 02: Cross correlation of the signals between original and extracted audio files.
Test 03: Format and characteristics of the original and extract image.
Test 04: Content and the properties of the original and extract files (other than image and audio).

The researchers have gotten assistance from open-source applications to accomplish the significant testing tasks. The xcorrSound tool package is a Quality Assurance work package [34]. The tool compares sound waves and is used in dissimilar digital audio conservation consequences. This entire tool package includes three distinct tools: overlap analysis, sound match, and waveform comparison. The researchers have used sound-match and waveform-compare to achieve the first two tests. DiffImgtool is a portable software tool that compares images of the same size as inputs [35]. So, Test 03 is covered by the DiffImgtool 2.0.0 version. Test 04 was done using the WinMerge software tool, which allows for the comparison of the content of the two files [36]. WinMerge 2.14.0 is a stable release, and this assists when a file or document is used as the hidden message rather than an audio or image file.

A survey was conducted to measure system performance among professionals to evaluate the quality of the system that satisfies the test areas among IT Professionals. The following Table II, illustrates the analyzed results of the test areas according to the test numbers.

TABLE II
Explore the results for Regular algorithm regarding to the test numbers.

| Test Number | Satisfaction | |
| --- | --- | --- |
| | Success | Fail |
| Test 01 | 93.1% | 6.9% |
| Test 02 | 94.2% | 5.8% |
| Test 03 | 97.1% | 2.9% |
| Test 04 | 99.3% | 0.7% |

The analysis of audio steganography performance reveals promising results across various types of embedded content. When considering audio properties of both the cover and embedded audio, the success rate is impressively high at 93.1%. This indicates that the steganographic method effectively preserves the auditory characteristics of the carrier file while successfully hiding the embedded audio. The remaining 6.9% failure rate can be attributed to various factors, with noise in the embedded audio playing a significant role in these instances.

The study further explores the accuracy of extraction when using another audio file as the secret message (Test 02). In this scenario, the success rate slightly improves to 94.2%, with only 5.8% of cases failing to accurately extract the hidden audio. This marginal improvement suggests that the method is particularly well-suited for audio-in-audio steganography.

Interestingly, when images are used as the secret message, the performance differs noticeably from audio-based embedding. The application achieves a 97.1% success rate for image embedding, which is higher than both audio-based tests. This could be due to the different nature of image data and how it interacts with the steganographic algorithm. Most impressively, for other types of messages (possibly text or binary data), the success rate peaks at 99.3%. This exceptionally high performance for non-audio and non-image data types indicates that the steganographic method is particularly effective for these forms of content, possibly due to their more compact or structured nature compared to audio or image files.

TABLE III
Explore the results for Excessive algorithm regarding to the test numbers.

| Test Number | Satisfaction | |
| --- | --- | --- |
| | Success | Fail |
| Test 01 | 90.1% | 8.9% |
| Test 02 | 94.2% | 5.8% |
| Test 03 | 96.8% | 3.2% |
| Test 04 | 99.3% | 0.7% |

The excessive algorithm introduces an innovative approach to audio steganography, resulting in performance characteristics that differ from the regular algorithm (as shown in Table III). This method allocates additional space for the secret message, which leads to notable differences in the properties of both the cover audio and the embedded audio when compared to the conventional approach.

The success rate for this excessive algorithm experiences a slight decrease to 90.1%, primarily due to the increased influence of noise in the embedded audio. This reduction in success rate suggests that the additional space allocation may introduce more noticeable artifacts in the carrier audio. However, the exactness between the original and extracted audio file remains consistent with the regular algorithm, indicating that the extraction process maintains its efficiency despite the changes in the embedding method.

Interestingly, when using images as the secret message, the performance diverges from audio-based embedding. The application achieves a 96.8% success rate for image embedding, which is slightly lower than the regular algorithm but still impressively high. For other types of messages, the success rate remains at a peak of 99.3%, matching the performance of the regular algorithm. This consistency in handling non-audio and non-image data types suggests that the excessive algorithm maintains its effectiveness for these forms of content.

It's important to note that these percentages are derived from average results obtained through multiple tests. The evaluation process utilized various open-source tools, including xcorrSound, DiffImgtool 2.0.0, and WinMerge 2.14.0, along with basic properties directly accessible from the operating system [36]. This comprehensive testing approach ensures a thorough and reliable assessment of the algorithm's performance across different types of embedded content.

## 6. Discussion

This study explores an innovative approach to audio steganography, focusing on concealing messages within audio files. The significance of audio-based information hiding is multifaceted and potentially far-reaching, particularly in its ability to mitigate man-in-the-middle attacks. While certain threats, such as F5 attacks, pose challenges to steganographic methods, the algorithms developed in this research demonstrate robust capabilities to counter such threats [37].

These advanced algorithms surpass traditional LSB (Least Significant Bit) techniques by employing a more sophisticated method of data integration within the audio file. This approach not only provides accurate and enhanced security for data hiding but also incorporates a cryptographic layer using AES-256 bit encryption for the secret message. The algorithms are designed to maintain the quality of the carrier audio file post-embedding and ensure the accuracy of the extracted hidden information. Notably, while modern audio formats often exceed the traditional 8-bit or 16-bit structures found in formats like WAV, the proposed solution adapts by considering changes across 8-bit segments [38]. This approach allows for compatibility with various audio formats without compromising the quality of the cover object, demonstrating the versatility and effectiveness of the steganographic technique in contemporary digital audio environments.

## 7. Conclusion

The pursuit of a perfect security system for digital audio steganography remains an ongoing challenge, given the inherent limitations of realistic cover sources. However, the proposed technique in this study makes significant strides in enhancing the security and effectiveness of audio-based information hiding. This innovative approach combines a modified two-bit LSB (Least Significant Bit) substitution method with traditional LSB techniques for embedding secret messages. By utilizing both methods, the system achieves a balance between increased data capacity and maintaining imperceptibility. The evaluation of the system focused on two critical functionalities: successfully concealing the existence of the hidden message and accurately extracting it when required. A notable feature of this system is the integration of encryption alongside steganography, providing an additional layer of security. This dual-layer approach allows users to ensure the confidentiality of their message even if the steganographic technique is compromised. The system's capability to transmit audio files securely over a local area network further enhances its practical applicability in real-world scenarios.

The synergy between cryptography and steganography in this system creates an exceptionally robust security framework. Compared to existing systems, this application offers enhanced features, improved security, and faster processing speeds. The robustness of the algorithm ensures that the hidden information remains intact even when subjected to various audio processing operations. It's important to note that the system does have limitations, particularly regarding the size of the secret message in relation to the cover file. This constraint is a common challenge in steganographic techniques and requires careful consideration when selecting cover files. The survey results provided offer a comparative analysis of the entire process, highlighting the system's performance against other existing solutions. The uniqueness of these mixing algorithms lies in their novel approach to combining various techniques, resulting in a new proposition for the field of information security.

In conclusion, while achieving perfect security remains an aspirational goal, this research presents a significant advancement in audio steganography. The proposed algorithms, with their unique combination of techniques, offer a promising new direction in the ongoing effort to secure digital communications.

## Declarations

## Authors' contributions

Sachith Dassanayaka contributed to the experiment design, analysis, performed the data, and involved in drafting the manuscript.

## References

[1]. Bender, W., Gruhl, D., Morimoto, N., & Lu, A. (1996). Techniques for data hiding. IBM systems journal, 35(3.4), 313-336.

[2]. Zwicker, E., & Fastl, H. (2013). Psychoacoustics: Facts and models (Vol. 22). Springer Science & Business Media.

[3]. Eranga, D. M. S., & Weerasinghe, H. D. (2015, August). Audio transporters for unrevealed communication. In 2015 Fifteenth International Conference on Advances in ICT for Emerging Regions (ICTer) (pp. 267-267). IEEE.

[4]. Jayasinghe, J. T., & Dassanayaka, S. (2024). Detecting Deception: Employing Deep Neural Networks for Fraudulent Review Detection on Amazon. https://doi.org/10.21203/rs.3.rs-5214171/v1

[5]. Dassanayaka, S. (2025). Sound Conveyors for Stealthy Data Transmission. arXiv preprint arXiv: 2502.10984.

[6]. Dassanayaka, S., Swed, O., & Volchenkov, D. (2024). Mapping the russian internet troll network on twitter using a predictive model. arXiv preprint arXiv:2409.08305.

[7]. Swed, O., Dassanayaka, S., Volchenkov, D.:Keeping it authentic: the social footprint of the trolls' network. Social Network Analysis and Mining14(1), 38 (2024).

[8]. G. J. Simmons, The prisoners' problem and the subliminal channel, (Advances in Cryptology. Proc. of Crypto 83, 1984), pp.51-67.

[9]. F Djebbar, B Ayad, H Hamam, K Abed-Meraim, A view on latest audio steganography techniques, Innovations in Information Technology (IIT), 2011 International Conference on , vol., no., (Abu Dhabi, 25-27 April 2011), pp. 409–414

[10]. Fallahpour, M., & Megias, D. (2009). High capacity audio watermarking using FFT amplitude interpolation. IEICE Electronics Express, 6(14), 1057-1063.

[11]. F Djebbar, D Guerchi, K Abed-Maraim, H Hamam, Text hiding in high frequency components of speech spectrum, Information Sciences Signal Processing and their Applications (ISSPA), 2010 10th International Conference on , vol., no., (Malaysia, 10-13 May 2010), pp. 666–669

[12]. Djebbar, F., Ayad, B., Abed-Meraim, K., & Hamam, H. (2013). Unified phase and magnitude speech spectra data hiding algorithm. Security and Communication Networks, 6(8), 961-971.

[13]. Djebbar, F., Abed-Meraim, K., Guerchi, D., & Hamam, H. (2010, May). Dynamic energy based text-in-speech spectrum hiding using speech masking properties. In 2010 The 2nd International Conference on Industrial Mechatronics and Automation (Vol. 2, pp. 422-426). IEEE.

[14]. Djebbar, F., Hamam, H., Abed-Meraim, K., & Guerchi, D. (2010, October). Controlled distortion for high capacity data-in-speech spectrum steganography. In 2010 Sixth International Conference on Intelligent Information Hiding and Multimedia Signal Processing (pp. 212-215). IEEE.

[15]. Djebbar, F., Ayad, B., Meraim, K. A., & Hamam, H. (2012). Comparative study of digital audio steganography techniques. EURASIP Journal on Audio, Speech, and Music Processing, 2012, 1-16.

[16]. Geiser, B., & Vary, P. (2008, March). High rate data hiding in ACELP speech codecs. In 2008 IEEE International conference on acoustics, speech and signal processing (pp. 4005-4008). IEEE.

[17]. YF Huang, S Tang, J Yuan, Steganography in Inactive Frames of VoIP Streams Encoded by Source Codec. IEEE Trans. Inf. Forensics and Security. 6(2), 296–306 (2011)

[18]. Johnson, N. F., & Jajodia, S. (1998). Exploring steganography: Seeing the unseen. Computer, 31(2), 26-34.

[19]. Koshkina, N. V. (2013). Detection of hidden messages embedded in audio signals by Hide4PGP. Journal of automation and information sciences, 45(5).

[20]. Ru, X. M., Zhang, H. J., & Huang, X. (2005, August). Steganalysis of audio: Attacking the steghide. In 2005 International Conference on Machine Learning and Cybernetics (Vol. 7, pp. 3937-3942). IEEE.

[21]. Qiao, M., Sung, A. H., & Liu, Q. (2009, June). Steganalysis of mp3stego. In 2009 International Joint Conference on Neural Networks (pp. 2566-2571). IEEE.

[22]. Q Liu, AH Sung, M Qiao. (2009), Temporal derivative-based spectrum and mel-cepstrum audio steganalysis. IEEE Trans. Inf. Forensics and Security. 4(3), 359–368

[23]. N Cristianini, J Shawe-Taylor, An introduction to Support VectorMachines.(Cambridge University Press, 2000)

[24]. J Nafeesa Begum, K Kumar, DrV Sumathy. (2010), Design And Implementation Of Multilevel Access Control In Medical Image Transmission Using Symmetric Polynomial Based Audio Steganography. Int. J. Comput. Sci. Inf. Security. 7, 139–146

[25]. M Shirali-Shahreza, Steganography in MMS, IEEE International Conference in Multitopic, INMIC 2007, (Lahore, Pakistan, 2007), pp. 1–4

[26]. M Paik, Blacknoise: Low-fi Lightweight Steganography in Service of Free Speech, Proceedings of the 2nd International Conference on M4D-Mobile Communication Technology for Development, NYU. (Kampala, Uganda, November 2010), pp. 1–11

[27]. Marler, Peter (2004). Nature's Music: The Science of Birdsong. Academic Press Inc. p. 207. ISBN 978-0124730700.

[28]. Oppenheim, Alan V.; Schafer, Ronald W.; Buck., John R. (1989). Discrete-time signal processing, Volume 2. Englewood Cliffs: Prentice-hall.

[29]. Shannon, Claude E. (January 1949). Communication in the presence of noise. Proc. Institute of Radio Engineers. 37 (1): 10–21. Reprint as classic paper in: Proc. IEEE, Vol. 86, No. 2, (Feb 1998)

[30]. Niels Ferguson; Richard Schroeppel; Doug Whiting (2001). A simple algebraic representation of Rijndael. Proceedings of Selected Areas in Cryptography, 2001, Lecture Notes in Computer Science. Springer-Verlag, Archived from the original (PDF/PostScript), (4 November 2006), pp. 103–111.

[31].     Lin, E., Woertz, E., Kam, M., LSB steganalysis using support vector regression, Information Assurance Workshop, Proceedings from the Fifth Annual IEEE SMC, (2004), pp 95 – 100.

[32]. G. Swain and S.K. Lenka, "Steganography-Using a Double Substitution Cipher", International Journal of Wireless Communications and Networking, vol. 2, no. 1,(2010), pp. 35-39.

[33]. ALAN AGRESTI, Categorical Data Analysis, 2nd edition, (John Wiley & Sons, Gainesville, Florida, 2002).

[34]. Johnson, A. B. (2025, February 15). xcorrSound: waveform-compare New Audio Quality Assurance Tool. https://github.com/openpreserve/scape-xcorrsound

[35]. DiffImg Portable 2.0.0 (image comparison tool) Released, http://portableapps.com/news/2013-08-06--diffimg-portable-2.0.0-released

[36].  Dong, S., Abbas, K., & Jain, R. (2019). A survey on distributed denial of service (DDoS) attacks in SDN and cloud computing environments. IEEE Access, 7, 80813-80828.

[37]. Yan, Q., Yu, F. R., Gong, Q., & Li, J. (2015). Software-defined networking (SDN) and distributed denial of service (DDoS) attacks in cloud computing environments: A survey, some research issues, and challenges. *IEEE communications surveys & tutorials*, *18*(1), 602-622.

[38]. Fleischman, E. (1998). WAVE and AVI codec registries (No. rfc2361).